\documentclass[traditabstract]{aa} % for a referee version add ,referee
\usepackage{txfonts,graphicx}

\begin{document}

\title{Intrinsic scatter of caustic masses and hydrostatic bias: An
observational study.
}
\titlerunning{Intrinsic scatter of caustic masses and hydrostatic bias} 
\author{S. Andreon\inst{1} \and G. Trinchieri\inst{1} \and A. Moretti\inst{1}
\and J. Wang\inst{2} 
}

\authorrunning{Andreon et al.}
\institute{
$^1$INAF--Osservatorio Astronomico di Brera, via Brera 28, 20121, Milano, Italy
\email{stefano.andreon@brera.inaf.it}\\
$^2$Dep. of Physics and Astronomy, University of the Western Cape, Cape Town 7535, South Africa\\
}
\date{Accepted ... Received ...}
\abstract{
All estimates of cluster mass have some intrinsic scatter and perhaps some bias
with true mass even in the absence of
measurement errors for example caused by cluster triaxiality and large scale structure. 
Knowledge of the bias and scatter values is fundamental for both
cluster cosmology and astrophysics. In this paper we show that 
the intrinsic scatter of a mass proxy can be constrained by measurements of 
the gas fraction because
masses with higher values of intrinsic scatter with true mass 
produce more scattered gas fractions. Moreover, the relative bias of two mass estimates
can be constrained by comparing the mean gas fraction at the same (nominal)
cluster mass. Our observational study 
addresses the scatter between caustic (i.e., dynamically estimated) and true masses,
and the relative bias of caustic and hydrostatic masses. For these
purposes, we used the X-ray Unbiased Cluster Sample, 
a cluster sample selected independently from the intracluster
medium content with reliable masses: 
34 galaxy clusters in the nearby ($0.050<z<0.135$) Universe,
mostly with $14<\log M_{500}/M_\odot \lesssim 14.5$, and with caustic
masses. We found a 35\% scatter between
caustic and true masses.  Furthermore,
we found that the relative bias 
between caustic and hydrostatic masses is small, $0.06\pm0.05$ dex, improving
upon past measurements. The small scatter found confirms our previous
measurements of a highly variable amount of 
feedback from cluster to cluster, which is the cause of 
the observed large variety of core-excised X-ray luminosities
and gas masses.  
}
\maketitle

\section{Introduction}

Clusters of galaxies are the largest collapsed objects in the
hierarchy of cosmic structures (e.g., Sarazin 1988). 
They arise from the smooth sea of
hot particles and light under the action of gravity modulated
by the action of dark energy and dark matter 
(e.g., Dressler et al., 1996, Weinberg et al. 
2013). Our understanding of the gravitational processes that
shape the cosmic web, which allow us to use galaxy clusters as
cosmological probes (e.g., Vikhlinin et al. 2009),
and of the interplay between dark matter and baryonic components 
(e.g., Young et al. 2011 and references therein), relies
on scaling relations between halo mass and observable quantities tracing
one of their
constituting and observable parts, such as galaxies, 
intracluster medium, or dark matter.  All of the methods to weight galaxy clusters using these
observables
are subject to biases due to scatter between the mass and observable or
its dependency on other physical cluster properties. 
Even the direct observation of the total matter, via weak lensing,
is subject to scatter with mass due to cluster triaxiality, 
large scale structure, and
intrinsic alignments (Meneghetti et al. 2010; Becker \& Kravtsov 2011).

The caustic technique derives masses from measurements of the line-of-sight escape
velocity. Caustic masses are
unaffected by the dynamical state of the cluster and
by large scale structure (Diaferio 1999, Serra et al. 2011), however they are affected
by elongation along the line of sight. 
Previous analyses found that caustic masses have low scatter with true mass, but
the results are based on simulations, or are
indirect or noisy. In fact,
numerical simulations (Serra et al. 2011, Gifford \& Miller 2013) 
showed a 20\% scatter, 
Geller et al. (2013) showed that weak lensing and caustic masses agree within 30\%, and
Maughan et al. (2016) found a $23\pm11$\% scatter with hydrostatic masses.
Andreon \& Congdon (2014) found a small ($\ll 0.1$ dex) 
intrinsic scatter between richness 
and weak lensing mass, while
Andreon (2012) found a small ($\ll 0.1$ dex) scatter between 
richness and caustic masses.
Although indirect, the points above suggest that 
a large scatter between caustic and true
mass is unlikely, given the small scatter with weak lensing masses, 
hydrostatic masses,
and richness.  

The first aim of this paper is a data-driven determination of the intrinsic
scatter of caustic masses.
We achieve this objective by an innovative approach which can be applied
to other types of masses as well. We exploit the fact 
that a large scatter in halo mass
induces a large scatter on gas fraction (cosmic conspiracy notwithstanding);
the latter is proportional to one over halo mass.
Therefore, a value of scatter of the gas fraction can be
converted into an upper limit of the intrinsic scatter of the caustic masses
as detailed in Sect.~3. 
If caustic masses were low scatter proxies of true masses, then
they could be useful to calibrate noisier mass proxies. They could also be used  
to measure mass-related cluster properties
free of the large scatter of weak lensing masses due to
triaxiality, large scale structure, and intrinsic alignment.

A second aim is to measure the relative bias of caustic and hydrostatic masses. 
As mentioned, different methods
to estimate galaxy cluster masses
may also return systematically underestimated, or overestimated masses.
Hydrostatic masses, i.e., masses derived under the assumption of hydrostatic
equilibrium, have been often used both to measure intracluster medium 
properties (e.g., Vikhlinin et al.
2006; Arnaud et al. 2007) and to calibrate other mass proxies such as the
integrated pressure (e.g., Arnaud et al. 2010) and integrated pseudo-pressure (Arnaud et al. 2007,
Vikhlinin et al. 2009). Hydrostatic masses are known to be slightly 
biased estimates of true mass
because of deviations from the hydrostatic
equilibrium or the presence of nonthermal pressure support such as turbulence, bulk flows,
or cosmic rays (e.g., Rasia et al. 2006, Nagai et al. 2007, Nelson et al.
2014). The amount of the hydrostatic bias is uncertain, but usually estimated at 
10 to 20 \%. However, a much larger bias has been invoked to reconcile cosmological
parameters derived from the cosmic microwave background and cluster counts (Planck Collaboration 
2014), although the calibration of the bias by Planck team has been amply discussed (von der Linden
2014, Andreon 2014, Smith et al. 2016).
Caustic masses can provide an alternative calibration of the bias of hydrostatic masses.
They are almost unused for this purpose (we are only aware of  
Maughan et al. 2016) and we exploit a new method for using this type of masses: if there is a relative bias
between caustic and hydrostatic masses then there should be an offset in gas mass derived
using the two masses. 

Throughout this paper, we assume $\Omega_M=0.3$, $\Omega_\Lambda=0.7$, 
and $H_0=70$ km s$^{-1}$ Mpc$^{-1}$. 
Results of stochastic computations are given
in the form $x\pm y$, where $x$ and $y$ are 
the posterior mean and standard deviation. The latter also
corresponds to 68\% intervals because we only summarize
posteriors close to Gaussian in this way. Logarithms are in base 10.

\section{Sample selection, cluster masses, and X-ray data}

Sample selection, halo mass derivation, and X-ray data are presented and 
discussed in Andreon et al. (2016, Paper I), and gas masses are
derived in Andreon et al. (2017, Paper II), to which we refer for details. We summarize
here the work done. We used a sample of 34 clusters in the very nearby universe 
($0.050<z<0.135$, XUCS, for X-ray Unbiased Cluster Sample) 
extracted from the C4 catalog (Miller et al. 2005)
in regions of low Galactic absorption. 
There is no X-ray selection in our sample, meaning that 1) the probability of
inclusion of the cluster in the sample is independent of its X-ray luminosity
(or count rate), and 2) no cluster is kept or removed on the basis of its
X-ray properties, except for two clusters for which we cannot
derive gas masses. The impact of this selection is discussed in Sec.~3.3.
 
We collected the few X-ray observations present 
in the XMM-Newton or Chandra archives
and we observed the remaining clusters with Swift 
(individual exposure times 
between 9 and 31 ks), as detailed in Table 1 of Paper I. Swift observations have 
the advantage of a low X-ray background (Moretti et al. 2009), 
making it extremely useful for sampling a cluster
population that includes low surface brightness clusters
(Andreon \& Moretti 2011).  

Caustic masses within $r_{200}\footnote{The radius $r_\Delta$ is the
radius within which the enclosed average mass density is $\Delta$
times the critical density at the cluster redshift.}$, $M_{200}$, have been derived  following Diaferio \& Geller
(1997), Diaferio (1999), and Serra et al. (2011), then converted into $r_{500}$ and $M_{500}$
assuming a Navarro, Frenk \& White (1997) profile with concentration $c=5$. Adopting
$c=3$ would change mass estimates by a negligible amount; see Paper I. 
The median number of members within the caustics is $116$ and the
interquartile range is $45$. The median mass of the
cluster sample, $\log M_{500} / M_\odot$, is $14.2$ and the interquartile range is
0.4 dex. The average mass error is 0.14 dex.

Gas masses are derived by projecting a flexible radial
profile, fitting its projection to the unbinned X-ray data, and propagating
all modeled sources of uncertainties (e.g., spectral normalization, variation in exposure
time including those originated by vignetting or excised regions) 
with their non-Gaussian behavior (when relevant) into the gas mass estimate using
Bayesian methods. The spectral
normalization, measured in the annulus $0.15 < r/r_{500}<0.5$, is used to convert
brightnesses in gas densities. The average gas mass
error is 0.10 dex, as detailed and extensively tested, in Paper II.

\section{Analysis and results}

\begin{figure}
\centerline{
\includegraphics[height=7truecm]{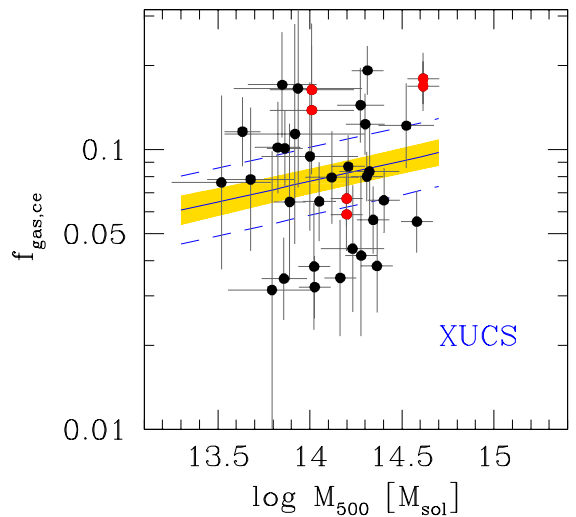}
}
\caption[h]{Core-excised gas fraction $f_{gas,ce}$ vs. halo mass $M_{500}$ of XUCS
sample. 
The solid line indicates the mean relations, the shading indicates its 68\% uncertainty,
whereas the dashed lines indicate the mean relation $\pm 1\sigma_{intr}$. 
Close pairs of red points (with identical masses) indicate measurements
from different X-ray telescopes.
}
\end{figure}

In Paper II, we fitted the relation between 
gas fraction and mass, allowing an
intrinsic scatter in $f_{gas}|M$,  while freezing the
scatter between caustic
and true masses to 0.08 dex. In this paper, we let it free to vary.

In detail, we  
allow caustic masses to have an additional scatter against true halo masses
$\sigma_{intr. caus.}$ to be added 
to our errors on $\log M^{obs}$, which already include a 20\% intrinsic
scatter between caustic masses and true masses (already accounted for in $\sigma_{\log M,i}$),
\begin{equation}
\log M^{obs}_{i} \sim N(\log M_i, \sigma^2_{\log M,i}+\sigma_{addit. scat.}^2) \ .
\end{equation}
We use 
a linear model with intrinsic scatter $\sigma_{intr}$ in (log) gas fractions,
\begin{equation}
\log M_{i,gas} \sim N( a (\log M_{i}-14)  + \log M_{gas,14} , \sigma_{intr}^2) \ . \\
\end{equation}
We fit the data in the gas mass versus halo mass plane, where errors are less
correlated (see Andreon 2010), i.e., 
\begin{equation}
\log M^{obs}_{i,gas} \sim N(\log M_{i,gas}, \sigma^2_{\log M_{gas},i})  \ . 
\end{equation}

Since we cannot properly determine the slope of the relation in the limited range
covered by XUCS clusters, we adopt as prior the posterior derived in 
Andreon (2010) for the sample in 
Vikhlinin et al. (2006) and Sun et al. (2009), $0.15\pm0.03$ as follows: 
\begin{equation}
a \sim N(1+0.15,0.03^2) .
\end{equation}
As shown in Sec.~3.3, the assumption of the slope value does not
change the results because the clusters studied have similar masses. 
For the remaining parameters (namely: 
additional caustic scatter, intrinsic scatter, and 
$\log$ of gas fraction at $\log M/M_\odot=14$, $\log f_{gas,14}=\log M_{gas,14}-14$), 
we assume a uniform and wide range of values that  
includes the true value as follows: 
\begin{equation}
\log M_{i}  \sim U(13.3,15.5)  
\end{equation}
\begin{equation}
\sigma_{intr} \sim U(0.01,1)  \\
\end{equation}
\begin{equation}
\sigma_{addit. scat.} \sim U(0,3)  \\
\end{equation}
\begin{equation}
\log M_{gas,14} \sim U(10,14) \ . \\
\end{equation}
The parameter space is sampled by Gibbs sampling using JAGS
(see Andreon 2011). For the three clusters for which multiple  
estimates for gas fraction are available,  derived
from different telescopes, the fit only uses those with smaller errors.

As detailed in Paper II, 
knowledge of the selection function in the observable and its 
covariance with the studied quantity is in general essential to propagate
selection effects from the quantity used to select the sample 
to the quantity of interest (here gas mass). The use of an
X-ray unbiased cluster sample, such that used in our paper,
does not require the application of any correction for 
the selection function to fit (often difficult to apply).

Using core excised gas masses, we found
\begin{equation}
\log f_{gas} =  (0.15\pm0.03)\  (\log M_{500}-14) -1.11\pm0.04 \ ,
\end{equation}
with the slope posterior
largely determined by the slope prior. 

The results of the fit are plotted in Fig.~1, including 
the mean relation (solid
line), its 68\% uncertainty (shading), and the mean
relation $\pm 1 \sigma_{intr}$.

\begin{figure}
\centerline{
\includegraphics[width=9truecm]{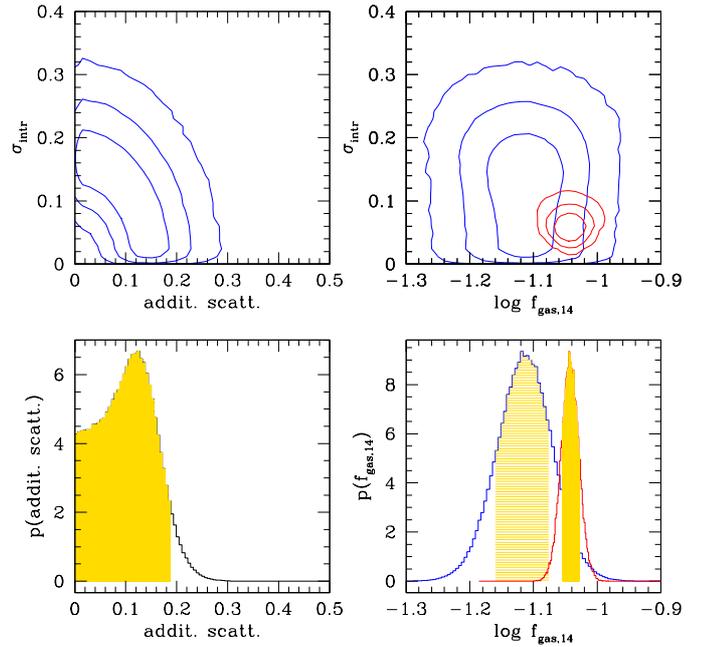}
}
\caption[h]{Joint 68\%, 95\%, and 99.7\% (highest posterior) probability
contours (top panels) of our fit to XUCS cluster (in blue) and to an
X-ray selected sample of relaxed clusters (in red, from Andreon 2010). 
The bottom panels show
the marginals:
the posterior probability distribution of the gas fraction (bottom right panel)
and the additional scatter of caustic masses (bottom left panel). The shading indicates
the 68 \% (highest posterior) probability in the bottom right panel and the 
95 \% probability in the bottom-left panel. 
} 
\end{figure}

\subsection{Constraints on the scatter of caustic masses from the gas fraction scatter}

The top left panel of Figure~2 shows the joint
posterior probabilities distributions of intrinsic and additional {\it scatters}.
The scatter of the data sets a joint constraint on the 
intrinsic scatter and on the additional caustic scatter (approximatively on their
sum). In particular, if $\sigma_{intr}\gtrsim 0.15$,  
the most probable additional scatter is $<0.1$, i.e., caustic masses have very little 
additional scatter with true mass above what has already been considered
in the mass error (0.08 dex).
If $\sigma_{intr}\approx0$  
then the most probable value of additional scatter is $\sim0.15$ dex, i.e.,
scatter with true mass is underestimated by 40\% at most. 
For whatever (positive) intrinsic scatter, the additional caustic scatter 
cannot exceed 0.3 dex (bottom left panel) 
because higher values require a data scatter that is  
larger than the one observed. 
Finally, marginalizing (averaging) over
all possible values, the additional caustic scatter is less than 0.19 dex 
with 95\% probability (see bottom left panel), which brings the posterior
mean of the total scatter to 0.13 dex and
the 95\% upper
limit of the total scatter of caustic masses to 0.21 dex. 
This observation-based 35\% scatter 
confirms and improves upon other indirect or
noisy evidence (see introduction) and 
makes caustic masses the prime choice to measure 
mass, in particular 
free of the biases of other mass estimates (such as
hydrostatic masses).   

The idea of exploiting the observed scatter in the gas fraction to derive an
upper limit to the scatter of a mass estimate can be
applied to other mass estimates, such as hydrostatic or weak lensing masses.
However, although the idea is promising, the application to other
samples needs to carefully account for complications which are absent in
our X-ray Unbiased Cluster Sample but present in
other samples such as: a) in samples selected using their content in gas mass (or a quantity
showing covariance with it, such as X-ray luminosity; see Paper II);
b) in samples including only 
a cluster subpopulation; c) in samples selected with unknown or ill-defined 
selection function, or d) in samples in which 
the reference radius used to determine the gas
fraction depends on the X-ray data (as in hydrostatic mass
estimates).

\begin{figure}
\centerline{
\includegraphics[width=7truecm]{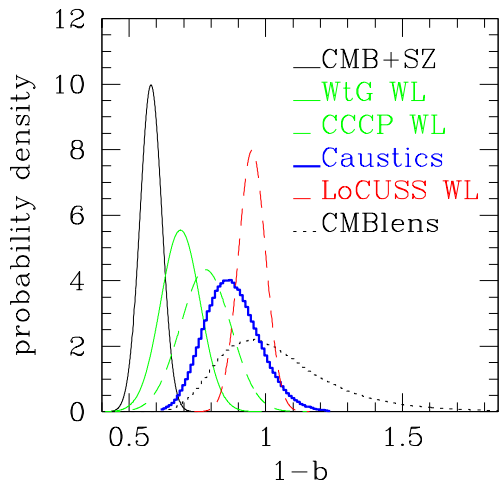}
}
\caption[h]{Comparison of estimates of the hydrostatic bias $1-b$. The leftmost (solid
black) curve 
is the distribution needed to reconcile cluster counts and cosmological
constraint from CMB (Planck collaboration XXIV 2016). The dotted broad distribution
reaching $1-b=1.5$ is the bias found using lensing of CMB
temperature anisotropies (Planck collaboration XXIV 2016). Other curves 
refer to weak lensing mass estimates (von der Linden et al. 2014; 
Hoekstra et al. 2015; Smith et al. 2016) or our work (blue thick curve
centered on $1-b=0.88\approx 1-0.06 \ln 10$). All probability densities integrate to one (as
probability axioms prescribe) and we converted without approximations 
the distribution in $1/(1-b)$ (from lensing from CMB)
to the plotted $1-b$ distribution.
}
\end{figure}

\subsection{The small relative bias of hydrostatic and caustic masses}

In this section we set a limit to the relative bias of hydrostatic 
and caustic masses by comparing the {\it average} gas fraction of clusters
with $\log M/M_\odot=14$, as derived using caustic
and hydrostatic masses. If a bias exists,
then it should appear as a difference in the average gas fraction
of clusters with the same nominal halo mass. The compared fractions are
based on XUCS for clusters with caustic masses and on  
Vikhlinin et al. (2006) and Sun et al. (2009) for clusters with
hydrostatic masses.

As mentioned in the introduction, a large
bias has been invoked to reconcile cosmological parameters
derived by the Planck team from different probes. Fig.~3 illustrates
the variety of the hydrostatic biases $1-b$ found; the 
lefmost solid line is the bias needed to reconcile cosmological parameters 
(Planck collaboration XXIV, 2016) and the other curves show independently derived
mass biases (von der Linden et al. 2014; Hoekstra et al. 2015; Smith et al.
2016), which are often called mass priors in the literature. As shown in the Figure,
there is some tension between the various determinations. However,
the selection function is either not available or not accounted for
in the computation of the bias (except for the black curve), which 
may affect the derived bias (see Andreon 2016).

The core-excised gas fraction vs. halo mass mean relation of XUCS has been
derived in Sect.~3.1, while the non-core-excised gas fractions were used for
clusters in Vikhlinin et al. (2006) 
and Sun et al. (2009) and the relation was derived in Andreon (2010). 
The bottom right and top right panels of Fig.~2 show the posterior distribution
and joint probability contours for both samples.
The gas fractions of the two samples at 
$\log M/M_\odot=14$ only 
differ by $0.06\pm0.05$ dex, where XUCS gas fraction is lower, after accounting for
the negligible (0.01 dex) difference between core-excised and non-core excised gas masses
for XUCS clusters (derived in Paper II). A similar
conclusion may be qualitatively derived from Fig.~4 of Paper II,
where we plotted the fit on the Vikhlinin et al. (2006) 
and Sun et al. (2009) sample on XUCS individual values.

The close agreement of the two gas fractions may have two origins. First, 
there is a negligible relative bias ($0.06\pm0.05$ dex) 
between caustic and hydrostatic mass scales (see Fig.~3).
Our result is roughly consistent with that of Maughan et al. (2016)
($-0.08\pm0.04$). Second, the agreement is the result of
a fine tuning between an hidrostatic mass bias and sample selection. 
The comparison sample,
drawn from clusters in Vikhlinin et al. (2006) and Sun et al. (2009),
is formed by clusters selected to be relaxed 
but has otherwise an unknown representativeness because the
sample has an unknown selection function. 
Nevertheless, these clusters are those used to calibrate the observable-mass 
relations used by Vikhlinin et al. (2009) to constrain cosmological parameters,
which are found to be in agreement with those based
on other probes, suggesting that 
these clusters do not provide biased scaling
relations. To have the small offset between
the derived gas fraction as result of selection effects
compensating real differences, one of the two following conditions are
requested. First, hydrostatic
masses would be underestimated and gas-rich (and photon-rich) would be
clusters preferentially discarded
in Vikhlinin et al. (2006) and Sun et al. (2009). 
However, there is no
reason why gas-rich clusters should be  preferentially discarded. Second,  
hydrostatic masses would be overestimated, but there is no 
evidence in the literature for that. These make
fine tuning an unlikely possibility.

We therefore conclude that a
relative bias between caustic and hydrostatic masses is small, 
if it exists at all, at least in the redshift and mass ranges
explored by our data, i.e., for clusters  
in the nearby ($0.050<z<0.135$) Universe,
mostly with $14<\log M_{500}/M_\odot \lesssim 14.5$. 
If the result
can be extrapolated to slightly more massive clusters at intermediate
distances (i.e., to the Planck cosmological sample), then
the source of the tension between cluster counts and CMB
cosmological parameters should be looked for somewhere else, for example
invoking a possible non-self-similar evolution of cluster scaling relations
(Andreon 2014) or a common bias for caustic and hydrostatic masses.

\subsection{Sensitivity analysis}

Our starting sample is formed by 34 clusters that form an X-ray unbiased sample.
However, two of these clusters are discarded in the course
of the analysis because of their X-ray properties.
We verified whether this alters the original properties of the
sample. We checked that the intercept of eq.~1 changes by 
less than 0.01 dex if we reintroduce the two clusters 
using a gas mass predicted from $L_X$ (Paper II), 
which implies that our conclusion on the relative bias of hydrostatic
and caustic masses (sec~3.2) is unaffected. 

Two objects are an unsufficient number to alter the scatter of the whole sample (34 objects)
population and therefore our constraint on the scatter of caustic masses
from the gas fraction scatter is robust.

Finally, our analysis of the gas fraction versus mass assumed
as slope prior the posterior of Andreon (2010)
based on Vikhlinin et al. (2006) 
and Sun et al. (2009) clusters because of the limited range in mass covered by
XUCS clusters. If we instead took a
uniform prior on the angle in order to allow  different
slopes, we found 
identical scatters and intercept and also
joint posterior distributions of the
key parameters (gas fraction, intrinsic scatter, and additional scatter
of caustic masses)
close to those depicted in Fig.~2.
This shows the robustness of our 
conclusions on assumptions about the slope. 

\subsection{Revisiting our previous papers}

In Paper I we kept the intrinsic scatter between true and
caustic massed fixed at the value of 0.08 dex and we 
found a 0.5 dex scatter in $L_X|M_c$, which is
a surprising large value not seen before. Smaller scatters were
probably the result of the selection of the samples through 
the ICM content. Paper I assumed, however, that caustic masses
have a small scatter with true mass. In the current paper
we show that our assumption was correct and therefore
our interpretation was correct. Quantitatively,
to induce a 0.5 dex scatter in X-ray luminosity at a given mass, 
caustic masses should have a
$\sigma(M_c|M_t)\sim0.6$ dex scatter, that is ruled out by a large margin
by the scatter in the gas fraction (sec.~3.2). 
Therefore, the big variance in core-excised X-ray luminosities found in Paper I 
is real
and not induced by an unaccounted scatter between true and 
caustic masses.

In Paper II, we fitted gas fraction as we do in this paper, but freezing
the scatter between true and caustic masses at the value of 0.08 dex.
The intercept we derive here allowing caustic masses to have
a free amount of intrinsic scatter is in full agreement with that
derived in Paper II. The intrinsic scatter in gas fraction derived
in the present paper depends on the prior on the intrinsic scatter
adopted for caustic masses because, as mentioned above, the data only offer
a constraint on the combination of the two. The value of intrinsic
scatter found here, $0.12\pm0.06$ dex, is marginally
lower than found in Paper II, $0.17\pm0.04$ because the difference has
been attributed to the scatter between real and caustic masses.
Although lower (by a statistical insignificant difference), 
the current scatter confirms our conclusion in Paper II of 
a highly variable, cluster-to-cluster,
amount of gas within $r_{500}$. This indicates that
the large scatter in gas fraction seen in Paper II
is not due to an unaccounted for scatter in 
caustic masses. Instead, it is due to a highly variable, cluster-to-cluster,
amount of gas within $r_{500}$. These differences are much larger than found in
the literature for subsamples of the whole cluster population. 

To summarize, the stringent upper limit to the scatter between caustic and true 
masses confirms that results in Paper I and II are not due to our assumption
of a small scatter between true and caustic masses.

\section{Conclusions}  

In this paper we introduced an innovative approch to derive an upper
limit to the scatter between mass and proxy and to estimate the relative
bias of two mass proxies. We applied it to 
caustic masses, and we note that it can be applied to other types 
of masses.

We used a sample of 34 clusters with caustic masses that we observed in X-ray  
and whose selection is, at a given
cluster mass, independent of the intracluster medium content (see Paper I). 
In Paper II we
derived gas masses by projecting a flexible radial
profile and fitting its projection to the unbinned X-ray data.
We also fit gas mass versus halo mass while 
keeping the intrinsic scatter between true and
caustic massed fixed at the value of 0.08 dex.
In this paper we modify the fitting model to include an additional
source of scatter, i.e. the scatter between true and caustic masses.
We then exploit the fact that the observed scatter in the gas fraction is
inflated by halo mass errors, and therefore an estimate of the former
sets an upper limit on the scatter of caustic mass
estimates. We found a 35\% scatter between 
caustic and true masses. 

Then, we set a limit to the relative bias between hydrostatic 
and caustic masses by comparing the {\it average} gas fraction of clusters
with $\log M/M_\odot=14$ derived using caustic
and hydrostatic masses.  We found
a small, if any, difference, $0.07\pm0.05$ dex, with caustic masses 
being larger and with the caveat that, as other works in
literature, the sample (part of it, in our case) has an
unknown representativeness. This amount of bias is, by a large margin,
too small to reconcile cosmological parameters derived by the Planck team from
different probes.

The small scatter measured between caustic and true masses confirms
our results in previous papers, which assumed the scatter to be small. 
We therefore confirm that the overall gas fraction is 
different from cluster
to cluster, indicating a variable amount of 
feedback from cluster to cluster. By ruling out a
large scatter between true and caustic masses 
we also confirm that the
large variety of core-excised X-ray luminosities 
observed in Paper I is due to scatter into the amount of gas, 
rather than to the
gas mass distribution within $r_{500}$.

{}

\end{document}